\documentclass[12pt]{iopart}

\usepackage{graphics}
\newcommand{\bea}{\begin{eqnarray}}
\newcommand{\eea}{\end{eqnarray}}
\newcommand{\be}{\begin{equation}}
\newcommand{\ee}{\end{equation}}

\def\be{\begin{eqnarray}}
\def\ee{\end{eqnarray}}
\def\bd{\begin{displaymath}}
\def\ed{\end{displaymath}}

\def\NP{Nucl. Phys. }
\def\PR{Phys. Rev. }
\def\PRL{Phys. Rev. Lett. }
\def\PL{Phys. Lett. }
\def\jpg{J. Phys. G: Nucl. Part. Phys. }

\begin{document}
\title{Simple parametrization of  $\alpha$-decay spectroscopic factor 
in $150\le A\le200$ region}
\author{G. Gangopadhyay}
%\footnote{\email{ggphy@caluniv.ac.in}}}
\address{Department of Physics, University of Calcutta,
92 Acharya Prafulla Chandra Road, Kolkata-700 009, India}
\date{}

\begin{abstract}
Half life values for $\alpha$-decay in the mass region $A=150-200$ have been 
calculated in the Microscopic Super Asymmetric Fission Model. The interaction between the
$\alpha$-particle and the daughter nucleus has been obtained in the double 
folding model with the microscopic interaction DDM3Y1. Theoretical densities 
for the nuclei involved have been calculated in Relativistic Mean Field 
approach with the Lagrangian density FSU Gold. Spectroscopic factors for 
$\alpha$-decay, calculated as the ratios of the theoretically calculated and 
experimentally measured half life values, show a simple dependence on the mass 
number and the product of the number of valence protons and neutrons. The 
implications of the parameterization have been discussed. Finally, based on 
this simple relation, $\alpha$-decay half life values in a number of nuclei 
have been predicted.
\end{abstract}


\maketitle
%\keywords{Relativistic Mean Field, $150\le A\le200$, Alpha decay, Spectroscopic factor,
%$N_pN_n$ scheme}

\section{Introduction}

Studies on $\alpha$-decay have often been undertaken in heavy and superheavy 
mass regions to investigate the nuclear baryon density profile. Both $\alpha$-decay  
and cluster decay are known to take place through tunnelling of 
the potential barrier which may theoretically be constructed from the 
densities of the daughter nucleus and the lighter decay particle using
a suitable interaction. The analytical super asymmetric fission model was 
developed to describe this process\cite{ASAFM}.  In the present work, we 
employ the Microscopic Super Asymmetric Fission 
Model\cite{MSAFM} to calculate the tunnelling probability in the WKB approximation.
The potential between the $\alpha$-particle and the daughter 
nucleus is obtained microscopically in the double folding model by folding the proton and
neutron densities 
in the $\alpha$-particle and the daughter nucleus with some suitable interaction. 
The densities may be obtained either from some phenomenological recipe or from
a theoretical calculation. In the present work we utilize the second approach 
and use the microscopic densities obtained from Relativistic Mean Field (RMF) 
theory. This  method has the advantage that it can be extended to nuclei far 
from the stability valley where phenomenological densities are not applicable.

In a microscopic approach, the spectroscopic factor is the ratio of the
experimental decay constant and the decay constant calculated in a simple 
Gamow picture. 
The experimental decay constant is given by
\be \lambda=\nu S P\ee
where $\nu$ is the assault frequency, $P$ is the tunnelling probability and
$S$ is the spectroscopic factor. If we consider the $\alpha$ particle to be 
preformed at the surface, the theoretical results may be obtained by the 
product of the first two terms. 
Thus the spectroscopic factor in 
$\alpha$-decay  was introduced to incorporate
the preformation probability of $\alpha$-particle. It may be considered as the 
overlap between the 
actual ground state configurations of the parent and the configuration 
described by one $\alpha$-particle coupled to the ground state of the daughter.
In this interpretation, it is expected to be less than unity
as the probability of $\alpha$-particle already preformed at the nuclear
surface is 
small. It has been shown\cite{cl} in cluster decay that  for a daughter 
nucleus, close to the magic number, the spectroscopic factor scales as 
$S_\alpha^{(A-1)/3}$ where $A$ is the mass
of the cluster and $S_\alpha$ is the spectroscopic factor for alpha decay.  

The spectroscopic factor for $\alpha$-decay is seen to depend on the odd-even 
effect as its value for  an even-even daughter is seen to be nearly double than 
that for an odd-mass daughter\cite{cl}.
In nuclei spread over a large mass 
region, it should incorporate other nuclear structure effects like 
deformation, shell closure, odd-even effects, etc. 
It is useful to parametrize the spectroscopic factor in terms of various simple 
quantities. Such parametrizations were obtained for 
various other nuclear 
quantities in terms of suitable functions of valence neutron and proton 
numbers ($N_n$ and $N_p$, respectively)\cite{Casten1}. Particularly, we find 
that quantities such as B(E2) values, which are measures of deformation can be 
parametrized easily in the $N_pN_n$ scheme   We have already studied 
$\alpha$-decay in heavy nuclei\cite{plb2} and have have seen that the 
$\alpha$-decay spectroscopic factor varies smoothly with  such a suitable 
function. In the present work, we extend our work on $\alpha$-decay to a lighter 
mass region, {\em i.e.} between $A=150$ and 200. Our aim is to obtain a simple 
phenomenological expression which can predict the spectroscopic factor in this 
mass region.

\section{Theory}

RMF is now a standard approach in low energy nuclear structure. It can describe 
various features of stable and exotic nuclei including ground state binding 
energy, shape, size, properties of excited states, single particle structure, 
features of exotic nuclei, etc\cite{Ring}. Being based on Dirac phenomenology, 
it naturally includes the spin degrees of freedom. As mass regions far from
the stability valley show indications of spin-orbit quenching, RMF is 
extremely suited for investigation of these nuclei. There  are different 
variations of the Lagrangian density and also a number of different 
parameterizations in RMF. We employ a recently proposed Lagrangian density
\cite{prl}, FSU Gold, which involves self-coupling of the vector-isoscalar 
meson as well as coupling between the vector-isoscalar meson and the 
vector-isovector meson. This Lagrangian density has earlier been employed to 
obtain the proton nucleus interaction to successfully calculate the half life 
for proton radioactivity\cite{plb} as well as our earlier works on $\alpha$ and 
cluster radioactivity\cite{plb2,282,cluster}. In this work also, we have 
employed FSU Gold. 

Since the nuclear proton and neutron densities as a function of radius are of importance in our 
calculation, the equations have been solved in co-ordinate space. The strength 
of the zero range pairing force is taken as 300 MeV-fm for both protons and 
neutrons. We have assumed spherical symmetry. The microscopic density dependent 
M3Y interaction, obtained from a finite range nucleon-nucleon interaction by 
introducing a factor dependent on the nuclear density, is a class of interactions that are 
frequently used in calculating nucleus-nucleus interaction. In the present work,
we have employed the interaction DDM3Y1 which has an exponential density 
dependence
\bea
v(r,\rho_1,\rho_2,E)=C(1+\alpha\exp{(-\beta(\rho_1+\rho_2)}))
\times(1-0.002E)u^{M3Y}(r)\eea
used in Ref. \cite{Khoa} to study $\alpha$-nucleus scattering with the standard 
parameter values, {\em viz.} $C=0.2845$, $\alpha=3.6391$ and $\beta=2.9605$ 
fm$^3$. Here $\rho_1$ and $\rho_2$ are the densities of 
the $\alpha$-particle and the daughter nucleus, respectively and $E$ is the 
energy per nucleon of the $\alpha$-particle in MeV. DDM3Y1 uses the direct
M3Y potential $u^{M3Y}(r)$ based on the $G$-matrix elements of the
Reid\cite{Reid} NN potential. The weak energy dependence was 
introduced\cite{Khoa1} to reproduce the empirical energy dependence of the 
optical potential. The NN interaction has been  folded 
with the theoretical densities of $\alpha$-particle and the daughter nucleus
in their ground states using the code DFPOT\cite{dfpot} to obtain the 
potential between them. The barrier tunnelling 
probability for the $\alpha$-particle has been calculated in the WKB 
approximation. The assault frequency has been calculated from the decay energy 
following Gambhir {\em et al}\cite{gambhir}. 
 
In the WKB approximation, the $Q$-value occurs in the exponential. Consequently,
a small change in $Q$-value can lead to an order of magnitude change in the 
estimates of half life and theoretically calculated $Q$-values do not achieve 
the desired high accuracy. Following the usual practice, the $Q$-values (and 
the decay energies) have been taken from experiment and are from Ref. 
\cite{mass}. Unless later measurements are available, the experimental partial 
half life values for $\alpha$-decay have been computed from the compilation by 
Akovali\cite{comp}. The newer measurements are taken from the NNDC website
\cite{NNDC}.
 
\section{Results}
The value of spectroscopic factors are generally expected to be  
less than unity as the ground state of the parent contain 
contributions from many other configurations other than the one with the 
$\alpha$-particle already formed inside the nucleus and coexisting with the 
daughter in its ground state. We calculate its value as the ratio of the 
calculated half life to the experimentally observed value. The logarithms of 
spectroscopic factors are presented in columns marked `I' in Table \ref{S}. 
\begin{figure}[b]
\resizebox{\columnwidth}{!}{\includegraphics{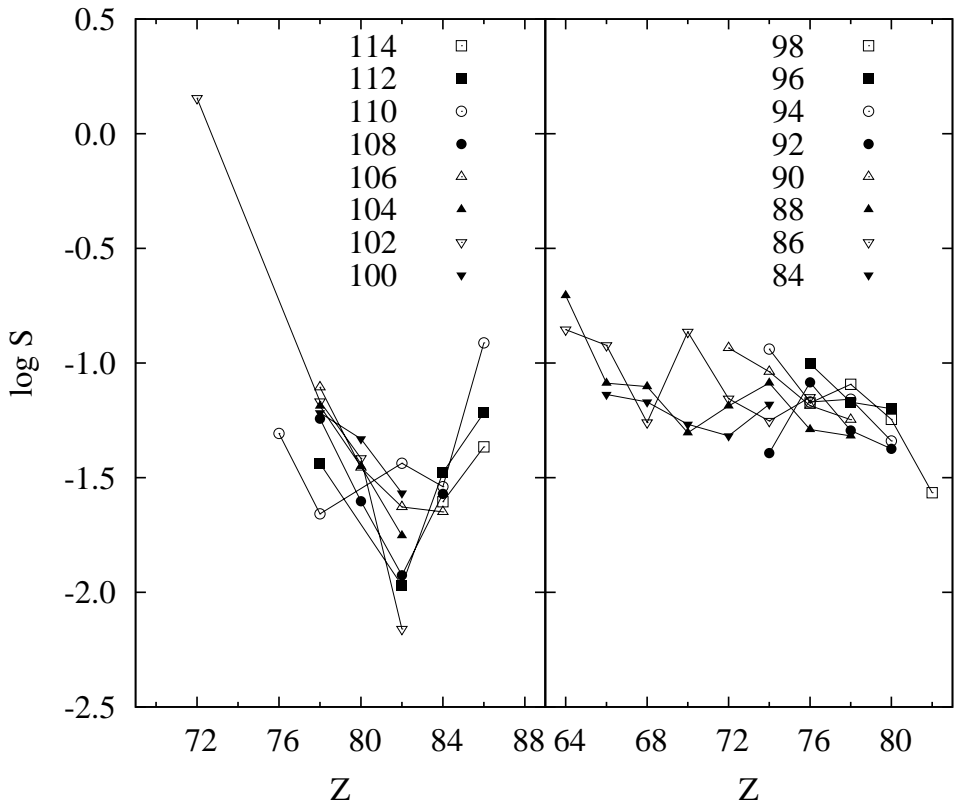}}
\caption{Spectroscopic factors for $\alpha$-decay as a function of parent proton number. See text for details.
\label{fig0}}
\end{figure}
\begin{table}[p]
\caption{Spectroscopic factors ($S$) for $\alpha$-decay. See text for details.\label{S}}
\begin{tabular}{ccccccccc}
\hline
Parent & N & \multicolumn{2}{c}{log($S$)}&
& N & \multicolumn{2}{c}{log($S$)}
\\
(Z)&&I& II&&&I& II\\\hline
Rn(86)&  114&  -1.366&  -1.479&
&  112&  -1.217&  -1.397\\
&  110&  -0.912&  -1.316\\
Po(84)&  116&  -1.706&  -1.699&
&  114&  -1.605&  -1.649\\
&  112&  -1.477&  -1.599&
&  110&  -1.539&  -1.549\\
&  108&  -1.571&  -1.499&
&  106&  -1.649&  -1.449\\
Pb(82)&  112&  -1.970&  -1.800&
&  110&  -1.437&  -1.782\\
&  108&  -1.926&  -1.763&
&  106&  -1.628&  -1.745\\
&  104&  -1.753&  -1.726&
&  102&  -2.160&  -1.708\\
&  100&  -1.567&  -1.689&
&   98&  -1.566&  -1.670\\
Hg(80)&  108&  -1.603&  -1.462&
&  106&  -1.456&  -1.412\\
&  104&  -1.451&  -1.362&
&  102&  -1.416&  -1.375\\
&  100&  -1.331&  -1.387&
&   98&  -1.246&  -1.400\\
&   96&  -1.199&  -1.413&
&   94&  -1.340&  -1.426\\
&   92&  -1.374&  -1.439\\
Pt(78)&  112&  -1.438&  -1.323&
&  110&  -1.658&  -1.242\\
&  108&  -1.243&  -1.160&
&  106&  -1.106&  -1.079\\
&  104&  -1.188&  -0.997&
&  102&  -1.168&  -1.042\\
&  100&  -1.217&  -1.086&
&   98&  -1.093&  -1.130\\
&   96&  -1.171&  -1.175&
&   94&  -1.158&  -1.219\\
&   92&  -1.294&  -1.263&
&   90&  -1.247&  -1.308\\
&   88&  -1.318&  -1.352\\
Os(76)&  110&  -1.307&  -0.972&
&   98&  -1.174&  -0.860\\
&   96&  -1.003&  -0.936&
&   94&  -1.170&  -1.012\\
&   92&  -1.084&  -1.087&
&   90&  -1.186&  -1.163\\
&   88&  -1.290&  -1.239&
&   86&  -1.149&  -1.315\\
W(74)&   94&  -0.939&  -0.804&
&   92&  -1.393&  -0.912\\
&   90&  -1.038&  -1.019&
&   88&  -1.087&  -1.126\\
&   86&  -1.253&  -1.233&
&   84&  -1.180&  -1.340\\
Hf(72)&  102&   0.155&  -0.043&
&   90&  -0.934&  -0.875\\
&   88&  -1.187&  -1.013&
&   86&  -1.156&  -1.152\\
&   84&  -1.318&  -1.290\\
Yb(70)&   88&  -1.304&  -0.900&
&   86&  -0.864&  -1.070\\
&   84&  -1.267&  -1.240\\
Er(68)&   88&  -1.103&  -0.787&
&   86&  -1.259&  -0.989\\
&   84&  -1.170&  -1.190\\
Dy(66)&   88&  -1.088&  -0.675&
&   86&  -0.923&  -0.908\\
&   84&  -1.137&  -1.140\\
Gd(64)&   88&  -0.706&  -0.750& 
&   86&  -0.854&  -0.952\\
\hline
\end{tabular}
\end{table}

The fact that the spectroscopic factors incorporates the effect of structure 
can be seen from the effect of shell closure. For example, in Fig. \ref{fig0} we 
plot the $ \log_{10}S$ as a function of parent  proton number for different 
isotones. It is clearly seen that despite the fluctuation in the value, except 
in the case of $N=110$, in all the nuclei up to $N=100$, $\log_{10}S$ shows a 
drop. In lighter nuclei, though the values for the shell closure are not
available in  most cases, we see that the values tend to decrease as one moves 
towards $Z=82$. Spectroscopic factors do tend to decrease in general as one 
moves from high $N$ values towards $N=82$, but the effect is less prominent.

\begin{figure}[b]
\resizebox{\columnwidth}{!}{\includegraphics{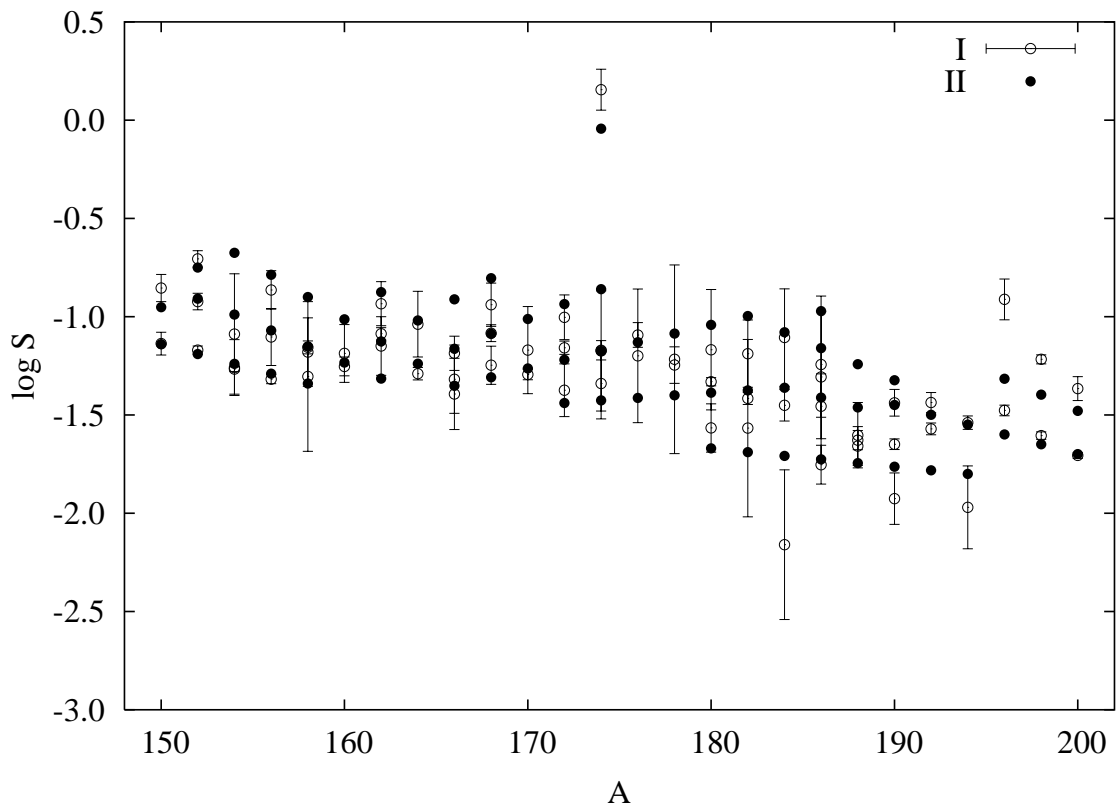}}
\caption{Spectroscopic factors for $\alpha$-decay. Here I and II refer to the
corresponding values in Table \ref{S}.\label{fig1}}
\end{figure}
Besides the effect of shell closure, the spectroscopic factors in the mass 
region show a weak downward trend with
mass number. There are differences for the same mass number 
corresponding to different $Z$ (and $N$) values. In an earlier calculation, we 
observed that the effect of different $Z$ and $N$ values may be expressed
as a simple function of number of  valence protons and neutrons\cite{plb}. This should also take care of the dependence on shell closure. In 
the present instance we find that the results for the logarithm of the 
spectroscopic factors can be very easily expressed as a simple function 
\be \log_{10}S_p=aA+bN_pN_n\ee
where $A$ is the mass number of the parent and $N_p$ and $N_n$ are the number 
of valence protons and neutrons, respectively. The number of valence particles
has been calculated using $Z=50$ and 82 as proton, and $N=82$ and
126 as neutron magic numbers. For the nuclei studied in the present work, the 
$\chi^2$ for the fit is 308. However, more than half the contribution comes 
from just six nuclei, $^{196,198}$Rn, $^{196,190}$Po, $^{192}$Pb,
and $^{188}$Pt. The parameters have thus been extracted without taking these 
nuclei in to account leading to a $\chi^2$ value of 99.0 for 63 nuclei. 
They are $a=-0.00928(8)$ and $b=0.00786(37)$. The values for $\log_{10}S_p$, 
calculated with the fitted parameters, are tabulated in Table \ref{S} in the 
columns marked `II'. We also plot in Fig. \ref{fig1}, the logarithm of the 
spectroscopic factors ($S$) calculated using RMF with errors and the values of 
$S_p$ obtained from eqn. (3) with the fitted parameters.  The errors include 
contributions from uncertainties in half life measurements and branching ratios 
as well as errors in the theoretical values due to the uncertainty in 
$Q_\alpha$ value. 

We should mention that of the six nuclei which had to be neglected in the 
fitting procedure, the adopted value for $^{196}$Po may actually have a larger 
error than estimated. The branching ratio to $\alpha$-decay, 98\%, is only an 
approximate value and may actually be substantially different. For example, 
Wauters {\em et al}\cite{Wauters} have measured it to be 94(5)\%. 
Thus it is possible that the estimated error for the $\alpha$-decay partial 
half life has a much larger error.

What is the significance of the dependence on $N_pN_n$? 
We have already mentioned that simplified parametrization of various nuclear 
quantities may be obtained if the quantities are plotted as functions of 
$N_p$ and $N_n$. We have, in an earlier work, shown that the spectroscopic 
factor increases with the Casten factor $N_pN_n/(N_p+N_n)$ \cite{plb}. 
Basically the product $N_pN_n$ is related to the integrated n-p interaction 
strength. Away from a closed shell nucleus, the integrated  strength increases 
and is parametrized by the product $N_pN_n$. Clearly this also incorporates the shell effects through the number of valence particles.  A positive value of the 
parameter `$b$' is thus consistent with our earlier 
observation in the mass region $A>208$. Essentially this also includes the 
effect of deformation, a degree of freedom not included in the present 
RMF calculation. The fact that the $N_pN_n$ scheme can take deformation into 
account has been observed in many instances\cite{Casten1}.

How may one interpret the negative value for the parameter `$a$'? As the mass 
number increases, the number of possible configurations contributing to the 
ground state also increases. Thus we may consider the contribution of a 
particular configuration, in this case that of the $\alpha$-daughter 
configuration, to the ground state of the parent nucleus should tend to 
decrease if all other effects are taken into consideration. In the present 
calculation, the effects of the shell structure and deformation have been 
incorporated in the factor $N_pN_n$. The negative value of the coefficient of 
$A$ expresses the above mentioned decreasing contribution of the 
$\alpha$-daughter configuration. 

There are several possible $\alpha$-decaying nuclei in this mass region for 
which the half life values have not yet been measured though the Q-values are 
known from mass measurements. We have theoretically calculated the half life
in our approach using the spectroscopic values from eqn. (3) with the fitted
parameters. We present our results in Table \ref{T}. We restrict ourselves to
the nuclei whose half life values are calculated to be less than 10$^{18}$
years. Most of these nuclei undergo beta-decay. We have also tabulated the 
calculated branching ratios, if beta-decay half life is known and 
the value is at least $10^{-6}$.
Because of the large errors in the Q-values, the half life in most of these 
nuclei may be wrong by a factor of two.
\begin{table}[h]
\caption{Calculated $\alpha$-decay partial half life values and branching 
ratios for a number of nuclei with positive $Q_\alpha$ values. See text for 
details.
\label{T}}
\begin{tabular}{lcll}\hline
Nucleus & $Q_\alpha$ (MeV) & \multicolumn{1}{c}{T$_{1/2}$} & $b_\alpha$(\%)\\\hline
$^{200}$Pb & 3.158 & 1.1$\times10^{16}$Yr\\
$^{198}$Pb & 3.718 & 1.4$\times10^{10}$Yr & \\
$^{196}$Pb & 4.225 & 7.5$\times10^5$Yr  &\\% 9.4$\times10^{-9}$\\
$^{192}$Hg & 3.387 & 8.0$\times10^{11}$Yr\\
$^{190}$Hg & 4.069 & 2.9$\times10^5$Yr\\
$^{188}$Po & 8.082 & 18.4 ms & 100?\\
$^{184}$Os & 2.963 & 1.3$\times10^{13}$Yr&100 \\
$^{182}$Os & 3.382 & 2.7$\times10^{8}$Yr\\
$^{180}$Os & 3.857 & 1.1$\times10^{4}$Yr\\
$^{180}$W & 2.508 & 1.8$\times10^{17}$Yr&100\\
$^{178}$Pb & 7.790 & 4.2 ms& 100?\\
$^{178}$Os & 4.256 & 14.5Yr & 6$\times10^{-5}$\\
$^{178}$W & 3.005 & 3.8$\times10^{10}$Yr&\\
%$^{178}$Hf & 2.080 & 1.2$\times10^{21}$Yr&\\
$^{176}$Os & 4.574 & 55 days & 5$\times10^{-5}$\\
$^{176}$W &  3.337 & 1.5$\times10^{7}$Yr&\\
$^{174}$W &  3.602 &  7.7$\times10^{4}$Yr&\\
$^{172}$W &  3.838 & 1.2$\times10^{3}$Yr&1$\times10^{-6}$\\
$^{172}$Hf & 2.746 &  1.7$\times10^{12}$Yr&\\
$^{170}$W & 4.141 & 8.9Yr & 8$\times10^{-5}$\\
$^{170}$Hf & 2.910 & 2.6$\times10^{10}$Yr&\\
$^{168}$Hf & 3.237 & 1.1$\times10^{7}$Yr&\\
$^{166}$Hf & 3.548 & 2.1$\times10^{4}$Yr&\\
$^{164}$Hf & 3.923 &  29Yr & 1$\times10^{-6}$\\
$^{164}$Yb &  2.611 &   6.5$\times10^{12}$Yr&\\
$^{162}$Yb &  3.047 &    1.2$\times10^{8}$Yr&\\
$^{160}$Yb &  3.618 & 4.0$\times10^{2}$Yr& 2$\times10^{-6}$\\
$^{158}$Er &  2.669  & 5.1$\times10^{10}$Yr&\\
$^{150}$Er &  2.296 & 1.6$\times10^{15}$Yr&\\
\hline
\end{tabular}
\end{table}

Very little information is available on the nuclei in Table \ref{T}. The two 
nuclei $^{188}$Po and  $^{178}$Pb have not yet been studied experimentally. 
However, the large 
Q-value, and hence the short half life, indicates that $\alpha$-emission will
be the dominant, if not the only, decay mode. Experiments suggest that 
the half life in $^{180}$W is 1.1$^{+8}_{-4}\times 10^{17}$year, a value
in agreement with our calculation. The branching ratio in $^{190}$Hg
has a upper limit $3.4\times10^{-7}$\% while our calculated value is 
1.3$\times10^{-8}$\%. A probable candidate for double beta decay,
$^{184}$Os is known to have a half life greater than 5.6$\times10^{13}$year.
Our calculation suggests a slightly smaller value. This prediction
may be relatively easier to verify than the other ones as  
$\alpha$-decay is expected to the dominant mode in this nucleus and 
the half life is comfortably within present measurement capabilities.

\section{Summary}

Theoretical valued for $\alpha$-decay half life in the mass region
$A=150-200$ have been calculated 
in the Microscopic Super Asymmetric Fission Model with the theoretical densities obtained 
from RMF calculations using DDM3Y1 microscopic interaction. Spectroscopic 
factors for $\alpha$-decay, calculated as the ratios of the 
theoretically calculated and experimentally measured half life values, show a 
simple dependence on the mass number and the product of the number of valence 
protons and neutrons, the latter being a measure of integrated n-p interaction
strength. The mass number dependence is related to the contribution of the 
daughter plus $\alpha$ particle configuration to the ground state of the parent.
Finally, based on this simple relation, we have predicted the
half life values for $\alpha$-decay in a number of nuclei.

\section*{Acknowledgment}

This work was carried out with financial assistance of the
Board of Research in Nuclear Sciences, Department of Atomic Energy (Sanction
No. 2005/37/7/BRNS). Discussion with Subinit Roy is gratefully acknowledged.

\end{document}